\documentclass{sigchi}

\doi{https://doi.org/10.1145/3313831.XXXXXXX}
\isbn{978-1-4503-6708-0/20/04}
\conferenceinfo{Submitted to UIST'20,}{October 20--23, 2020, Minneapolis, MN, USA}
\acmPrice{\$15.00}

\newif\ifshowcomments
\showcommentstrue
\ifshowcomments
\newcommand{\TODO}[1]{{\color{red}{[TODO: #1]}}}
\newcommand{\revised}[1]{{\color[rgb]{0.2,0.7,0.2}{#1}}}

\newcommand{\phil}[1]{{\color[rgb]{0.9,0.1,0.1}{#1}}}
\newcommand{\dc}[1]{{\color[rgb]{0.9,0.1,0.9}{D: #1}}}

\else
\newcommand{\TODO}[1]{}
\newcommand{\revised}[1]{}
\newcommand{\lzhu}[1]{}
\newcommand{\phil}[1]{}
\fi

\def\ie{\emph{i.e.}}
\def\eg{\emph{e.g.}}
\def\etal{{\em et al.}}

\usepackage{balance}       
\usepackage{graphics}      
\usepackage[T1]{fontenc}   
\usepackage{txfonts}
\usepackage{mathptmx}
\usepackage[pdflang={en-US},pdftex]{hyperref}
\usepackage{color}
\usepackage{booktabs}
\usepackage{textcomp}

\usepackage{microtype}        
\usepackage{ccicons}          

\usepackage{todonotes}
\usepackage{tabularx}
\usepackage{float}
\usepackage{blindtext}
\usepackage{amsmath}
\newcolumntype{C}[1]{>{\centering\arraybackslash}p{#1}}

\def\plaintitle{GrabAR: Occlusion-aware Grabbing Virtual Objects in AR}

\def\emptyauthor{}
\def\plainkeywords{Authors' choice; of terms; separated; by
  semicolons; include commas, within terms only; this section is required.}

\makeatletter
\def\url@leostyle{%
  \@ifundefined{selectfont}{
    \def\UrlFont{\sf}
  }{
    \def\UrlFont{\small\bf\ttfamily}
  }}
\makeatother
\def\pprw{8.5in}
\def\pprh{11in}

\definecolor{linkColor}{RGB}{6,125,233}
\hypersetup{%
  pdftitle={\plaintitle},
  pdfauthor={\emptyauthor},
  pdfkeywords={\plainkeywords},
  pdfdisplaydoctitle=true, 
  bookmarksnumbered,
  pdfstartview={FitH},
  colorlinks,
  citecolor=black,
  filecolor=black,
  linkcolor=black,
  urlcolor=linkColor,
  breaklinks=true,
  hypertexnames=false
}

\begin{document}

\title{\plaintitle}

\numberofauthors{1}
\author{Xiao~Tang,
	Xiaowei~Hu,
	Chi-Wing~Fu, 
	and~Daniel~Cohen-Or}

\teaser{
  \centering
  \includegraphics[width=\textwidth*\real{0.99}]{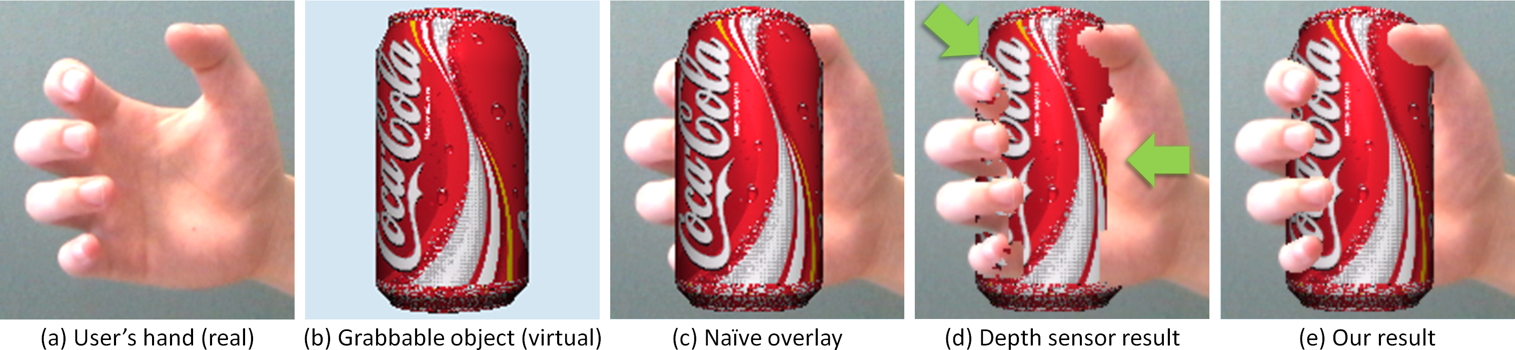}
  \vspace*{1mm}
  \caption{Most AR applications today ignore the occlusion between real (a) and virtual (b) objects when incorporating virtual objects in user's view (c).
  	Using depth from 3D sensors relieves the issue,
  	but the depth is not accurate enough and may not match the rendered depth for virtual objects, so undesirable artifacts often appear; see green arrows in (d).
  	Our GrabAR learns to compose real and virtual objects with natural partial occlusion (e).}
  \vspace*{-2mm}
  \label{fig:occlusion_problem}
}

\maketitle

\begin{abstract}
Existing augmented reality (AR) applications often ignore the occlusion between real hands and virtual objects when incorporating virtual objects in user's views.
The challenges come from the lack of accurate depth and mismatch between real and virtual depth.
This paper presents GrabAR, a new approach that {\em directly predicts the real-and-virtual occlusion\/} and bypasses the depth acquisition and inference.
Our goal is to enhance AR applications with interactions between hand (real) and grabbable objects (virtual).
With paired images of hand and object as inputs, we formulate a compact deep neural network that learns to generate the occlusion mask.
To train the network, we compile a large dataset, including synthetic data and real data.
We then embed the trained network in a prototyping AR system to support real-time grabbing of virtual objects.
Further, we demonstrate the performance of our method on various virtual objects, compare our method with others through two user studies, and showcase a rich variety of interaction scenarios, in which we can use bare hand to grab virtual objects and directly manipulate them.
\end{abstract}




\section{Introduction}
\label{sec:intro}



Augmented Reality (AR)~\cite{krueger1985videoplace,caudell1992augmented} has become more popular nowadays with many phone applications supporting it.
However, most AR applications today simply put virtual objects as an overlay above the real hands and cannot support free-hand interactions with the virtual objects.
It remains challenging to incorporate virtual objects into the real world, such that the real hands and virtual objects are {\em perceived to naturally co-exist\/} in the views, and further, we can {\em directly interact\/} with the virtual objects, as if they are in our physical world.


To achieve the goal, one main challenge is {\em handling the occlusion\/} between the real and virtual objects, since they may {\em fully\/} or {\em partially\/} occlude one another.
Figure~\ref{fig:occlusion_problem} shows an example: (a) user's hand in the AR view and (b) a virtual can.
If we simply draw the can over the hand (c), the result is unrealistic, since parts of the fingers should go above the can when the hand grabs the can.
Occlusion (or interposition) 
is a crucial visual cue~\cite{ono1988dynamic,anderson2003role}, allowing us to rank the relative proximity among objects in views.
It comes naturally in the real world, but is typically ignored in existing AR applications.

One common approach to handle the occlusion is to acquire the depth of our hand using 3D sensors, then determine the occlusion by comparing the depth of the hand and the rendered depth of the virtual objects~\cite{battisti2018seamless,feng2018resolving}.
Figure~\ref{fig:occlusion_problem}(d) shows a typical result.
Certainly, this depth-based approach alleviates the occlusion problem.
However, depth acquisition is often noisy and imprecise, so the hand-object boundary is usually erroneous with obvious flickering during the interactions.
Also, user's hand may easily penetrate the virtual object (see arrows in Figure~\ref{fig:occlusion_problem}(d)), since the real depth of the hand may not precisely match the virtual depth from renderings.
Particularly, without haptic feedback, it is hard for one to avoid penetration.
Furthermore, this approach requires an additional depth sensor attached and registered with the RGB camera.

\begin{figure*}[!t]
\centering
\includegraphics[width=\textwidth*\real{0.99}]{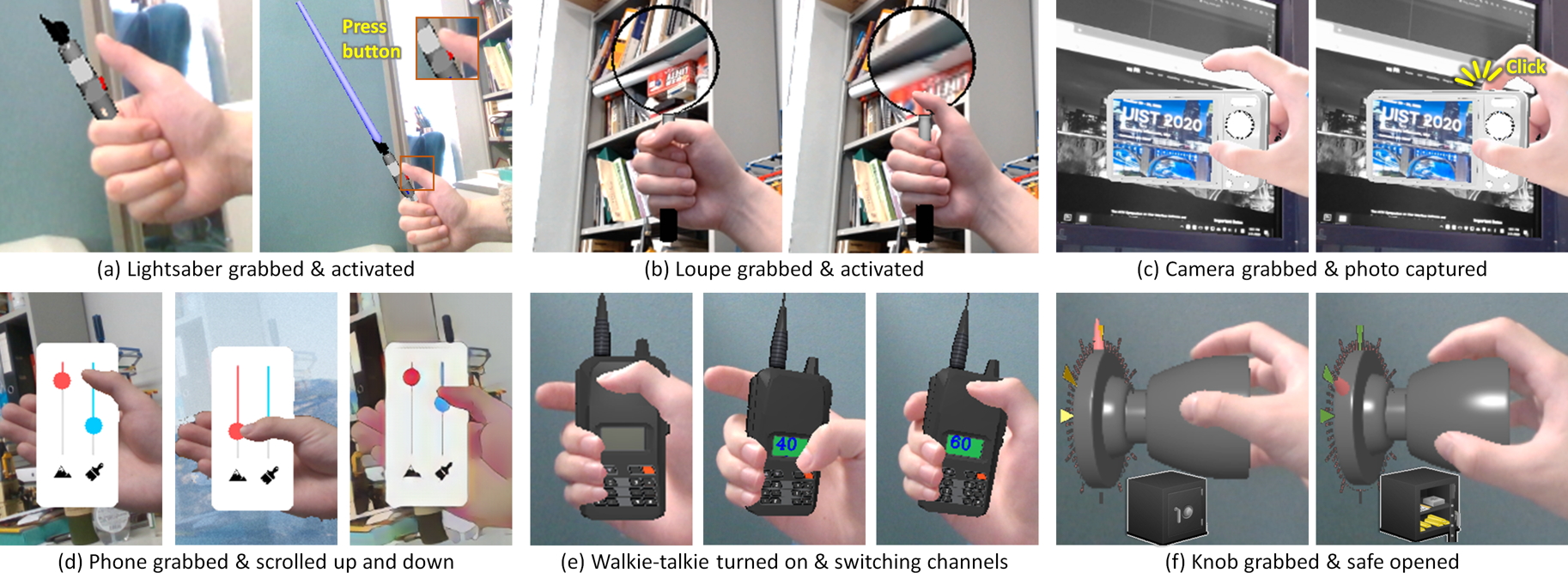}
\caption{Interaction scenarios empowered by GrabAR---user can grab these ``virtual'' objects and manipulate them interactively with functionality.}
\label{fig:scenarios}
\vspace*{-0.5mm}
\end{figure*}

Recently, RGB-based hand shape estimation has drawn attention~\cite{ge20193d,boukhayma20193d} with the potential
to solve the occlusion problem.
However, the approach has several drawbacks.
First, the training data for hand shape estimation is hard and tedious to collect, since annotations are difficult and time-consuming.
Second, existing datasets,~\eg,~\cite{Zimmermann2019ICCV}, mostly use a general hand model to represent the hand shape, so the real hand and hand model often misalign, leading to obvious artifacts in our views.
Third, hand shape estimation is a regression problem, which is sensitive to outliers; hence, the robustness of the existing methods is still weak.
We will provide visual comparisons with the state-of-the-art method later in the paper.

In this paper, we show that
\vspace*{-1mm}
\begin{quote}
	{\em by learning from natural grabbing poses in the image space,
	we are able to obtain plausible occlusion between the real hand and virtual objects, and further enable direct hand interactions with virtual objects.\/}
\end{quote}

In this work, we present
GrabAR, a new approach to resolve the occlusion between hand (real) and grabbable objects (virtual) by learning to determine the occlusion from the hand grabbing poses.
By using GrabAR, one can use his/her hand in physical space to grab the virtual objects in AR view, in which the hand can be naturally composed with the virtual object, {\em as if\/} the hand directly grabs a real object; see Figure~\ref{fig:occlusion_problem}(e).

Technically, GrabAR is a purely image-based approach that computes the occlusion entirely in the image space, without acquiring or inferring depth.
To design and train a deep neural network to determine the occlusion, the main challenges come from the time performance requirement and the dataset.
To this end, we design a {\em compact deep neural network\/} by adopting a {\em fast global context module\/} to aggregate global information of grabbing postures and formulating a {\em lightweight detail enhancement module\/} to enhance the boundary details between the real and virtual objects, with {\em two novel loss terms\/} to progressively focus on the regions of interest, while penalizing non-smooth boundaries.
%
Also, we compiled a large synthetic dataset and a small real dataset to enable effective network training.
%
%
Further, we implemented a prototyping AR system with the trained network and showcase a rich variety of interaction scenarios,~\eg, a scrollable virtual phone and a clickable virtual lightsaber, in which bare hand can directly grab and manipulate the virtual object, expanding the design space of AR applications; see Figure~\ref{fig:scenarios}.
Lastly, we performed various experiments both quantitatively and qualitatively, and two user studies, to evaluate our method among different methods.

\section{Related Work}

\begin{figure*}[t]
\centering
\includegraphics[width=\textwidth*\real{0.96}]{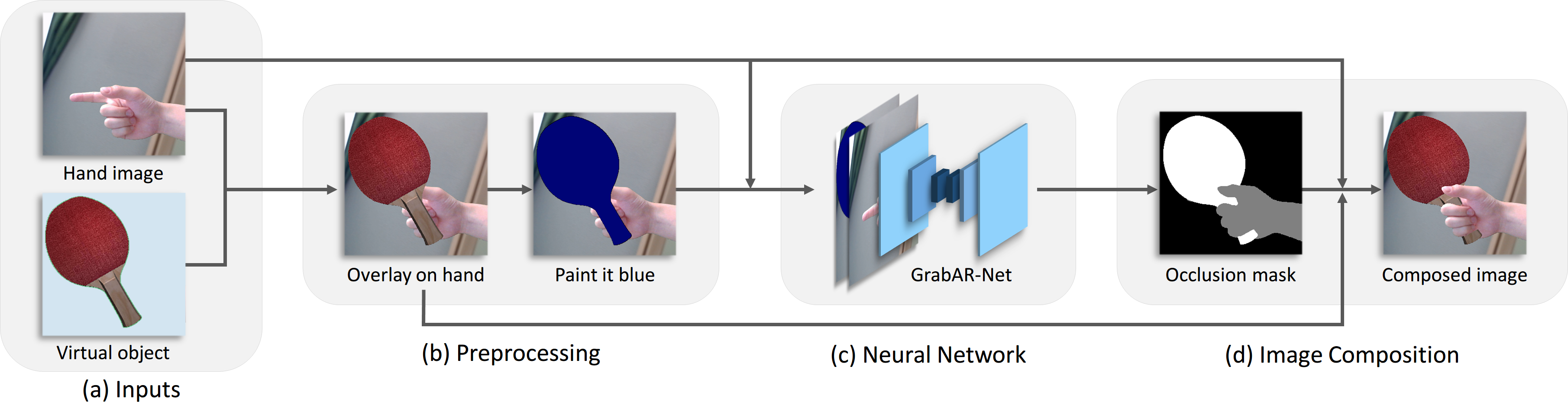}
\caption{GrabAR is a new approach to resolve the occlusion between real (hand) and virtual (grabbable) objects in the AR views.
Given an image pair of hand and virtual objects (a), we first preprocess them (b) and feed them to our neural network (c).
The network predicts an occlusion mask that indicates the hand regions that should overlay above the virtual object.
Hence, we can readily compose the two input images using the network-predicted occlusion mask, and produce a more natural composition of the real and virtual objects in the AR view (d).}
\label{fig:pipeline}
\vspace*{-1mm}
\end{figure*}


{\bf Hand-based AR interactions\/} were first realized using data gloves~\cite{dorfmuller2001finger,wang2009real,marquardt2011continuous},~\eg, Benko~\etal~\cite{benko2007balloon}.
%
%
However, the device is expensive, tedious to set up, fragile, and also, hinders the user actions.
%
Later, many works explore depth sensors and hand tracking for hand-object interactions in AR.
Radkowski~\etal~\cite{radkowski2012interactive} use the Kinect sensor to enable users to move virtual objects with fist/open hand gestures. 
%
%
Moser~\etal~\cite{moser2016calibration} implement an optical see-through AR system using LeapMotion~\cite{leapmotion}.
However, the depth sensor is used only to track the hand, without resolving the hand-object occlusion.

Some recent works begin to explore hand-object occlusions in AR.
%
Feng~\etal~\cite{feng2018resolving} use LeapMotion and combine a tracking-based method and a model-based method to estimate the occlusion mask.
Though some results are shown for few objects, their approach requires a cumbersome set up to customize user's hand shape on paper, then track and estimate the hand geometry using a depth sensor for producing the occlusion mask.
%
Battisti~\etal~\cite{battisti2018seamless} leverage stereo images from LeapMotion to obtain depth of hands, then compare it with rendered depth from virtual objects to determine the hand-object occlusion.
As discussed in the introduction, this approach, however, suffers from various issues due to depth inaccuracy and mismatch between the acquired depth and rendered depth.

\vspace*{-0.3mm}
Vision-based methods typically require a monocular camera to track user's hands, which are more generic for use,~\eg, with smartphones.
However, hand-pose and depth estimation in monocular views are still challenging, so most existing works focus on gesture recognition instead of occlusion resolution.
Chun~\etal~\cite{chun2013real} develop a simple hand recognition method for AR interactions.
Choi~\etal~\cite{choi2013ihand} estimate the six-DoF pose of a jazz hand for simple virtual object overlay.
Song~\etal~\cite{song2015joint} use an RGB camera to recognize hand shapes, estimate the mean hand-camera distance, and use the distance for assorted interactions, \eg, selection.
However, the result is still not sufficiently precise to resolve the hand-object occlusion.

{\bf Hand pose/shape estimation\/} aims to employ a 3D hand mesh to estimate an approximate 3D hand shape/pose from RGBD images~\cite{rogez2015first,mueller2017real,yuan2018depth,ye2018occlusion}, single RGB images~\cite{zimmermann2017learning,cai2018weakly,panteleris2018using,mueller2018ganerated,baek2019pushing,ge20193d,boukhayma20193d}, or single depth images~\cite{malik2018deephps}. 
Among them, Ge~\etal~\cite{ge20193d} and Boukhayma~\etal~\cite{boukhayma20193d} predict both the hand pose and 3D hand model.
However, their focus is on the hand shape/pose instead of precise depth for hand and fingers.
Hence, if we take their results to determine the occlusion relations, we still have the hand-object penetration problem, which is nontrivial on its own.
%
%
In this work, our new approach bypasses the depth estimation and avoid the hand-object penetration altogether by {\em directly\/} predicting the occlusion in the image space.


{\bf Occlusion estimation\/}~\cite{ren2006figure,hoiem2007recovering,teo2015fast,wang2016doc} is a long-standing problem,
aiming to estimate the object boundaries in images and to identify the occlusion relations between objects per boundary segment.
%
Early works extract edge features~\cite{ren2006figure} and geometric grouping cues~\cite{teo2015fast} to identify the foreground and background.
%
%
Recently, deep neural networks are explored to detect object boundaries and estimate the occlusion~\cite{wang2016doc,wang2018doobnet}, but 
%
so far, works in this area focus on foreground (figure) and background (ground).
We are not aware of any method that directly reasons the occlusion between real and virtual objects.




{\bf Single-view depth estimation}
and occlusion estimation are inter-related problems~\cite{wang2016doc}.
If we obtain accurate depth for an image, we can easily infer the occlusions.
Conversely, if we find the occlusion boundaries, we can take them to improve the depth estimation, as shown in works such as~\cite{ren2006figure,hoiem2007recovering}.




\vspace*{-0.3mm}
Nowadays, single-view depth estimation is usually powered by SLAM~\cite{engel2014lsd,mur2015orb}, which takes a video stream from a single camera as input and predicts depth by means of an optimization.
%
%
%
%
Valentin~\etal~\cite{google2018ar} use 6-DoF-pose trackers on smartphones to estimate a sparse depth map via stereo matching.
%
Holynski~\etal~\cite{holynski2018occlusion} leverage sparse SLAM points extracted from a video stream and depth edges predicted from optical flow.
However, SLAM-based methods typically assume static scenes, which may not hold in scenarios with hand interactions.

Another stream of work explores depth estimation for single image in monocular views, which benefits scenarios with dynamic interactions.
%
Early works learn to predict depth in images using hand-crafted features,~\eg, Markov random field~\cite{saxena2006learning,saxena2008make3d}, light flow~\cite{furukawa2017depth}, and perspective geometry~\cite{ladicky2014pulling}.
Recently, convolutional neural networks (CNN) are explored for predicting depth.
Eigen~\etal~\cite{eigen2014depth} adopt a CNN to predict a coarse depth map then apply another CNN to refine the result.
Laina~\etal~\cite{laina2016deeper} develop a deep residual network and leverage the reverse Huber loss to predict depth.
Recently, Fu~\etal~\cite{fu2018deep} formulate network learning as an ordinal regression problem and develop a spacing-increasing discretization strategy to discretize depth, whereas Nicodemou~\etal~\cite{nicodemou2018learning} adopt the hourglass network~\cite{newell2016stacked} to infer depth of human hand.

The occlusion relationship in our task can be inferred from depth estimation.
However, taking a depth estimation approach to the problem requires high-quality depth for hands, which is hard to acquire using commodity depth sensors.
%
More importantly, the estimated depth is often not accurate enough for resolving the hand-object penetration for grabbing gestures in AR.
%
Our GrabAR is a new approach that directly predicts the occlusion between real hand and virtual grabbable objects, and bypasses the need to predict or acquire depth.

\vspace{-0.75mm}
\section{Overview}
\label{sec:overview}


GrabAR enables not only a natural composition of real hand (camera view) and virtual object (rendering) with partial occlusion, but also direct use of bare hands in physical space to grab the virtual objects in AR view and manipulate them, to a certain extent.
Figure~\ref{fig:pipeline} presents an overview of our approach.
The input is a pair of hand (real) and object (virtual) images (a).
To compose them with partial occlusion, we first preprocess them (b) and feed the results into our deep neural network (c).
The network then predicts an occlusion mask that marks the portions of the hand over the virtual object.
Using this mask, we can then correctly compose the hand and virtual object in AR.
Note the partial occlusion achieved in our result (d).

The main challenges come from the dataset and the time performance requirement.
First, the real hand and virtual object images must be {\em compatible in the view space\/}, so the hand pose in training data should appear to grab the virtual object in the view.
Second, we need ground-truth occlusion masks to supervise the network training.
However, to obtain accurate ground truths is typically very tedious, due to the {\em manual labeling work\/}; yet, training a network often requires a {\em large amount of data\/}.
Lastly, the network inference should run in  {\em real-time\/}, so the network architecture cannot be too complex.

To meet these challenges, we first employ 3D renderings to produce a large synthetic data with automatically-generated labels.
Then, we compile a rather small set of real images with manual labels to overcome the domain gap of synthetic images.
So, we can leverage a large amount of training data and significantly reduce the burden of manual labeling.
Next, rather than directly taking the hand and virtual object images as the network inputs, we preprocess them by overlaying the virtual object on hand and coloring it in blue; see Figure~\ref{fig:pipeline} (b).
Hence, the virtual object can look more distinctive from the hand, for the network to better learn from the inputs.
Then, we formulate a compact neural network and train it to locate hand portions above virtual object (see ``Method'' Section).
Equipped with the network to generate occlusion masks, we further create a prototyping AR system and deliver various interaction scenarios.
Please watch the {\em supplementary video\/} for real-time capture of the various demonstrations.

\begin{figure}[!t]
	\centering
	\includegraphics[width=8.35cm]{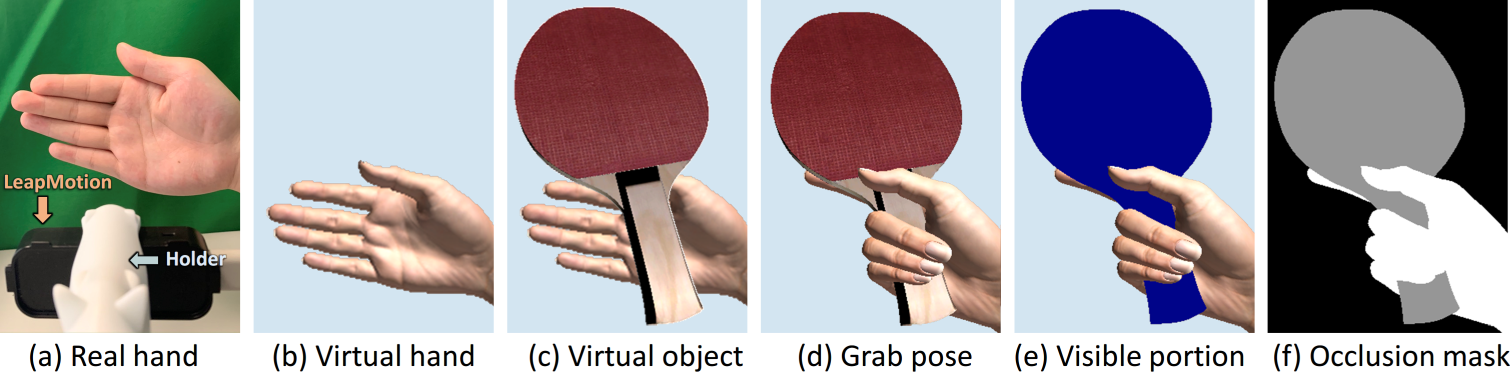}
	\caption{Synthetic data preparation.
		(a) Real hand in front of LeapMotion.
		(b) Rigged 3D hand model (virtual) corresponding to the real hand.
		(c) Target virtual object added next to the virtual hand.
		(d) Hand pose that appears to grab the virtual object.
		(e) Visible portions (blue) of the virtual object in the view.
		(f) Automatically-generated occlusion mask.}
	\label{fig:synthetic_data}
\end{figure}

\begin{figure}[!t] 
	\centering
	\includegraphics[width=8.35cm]{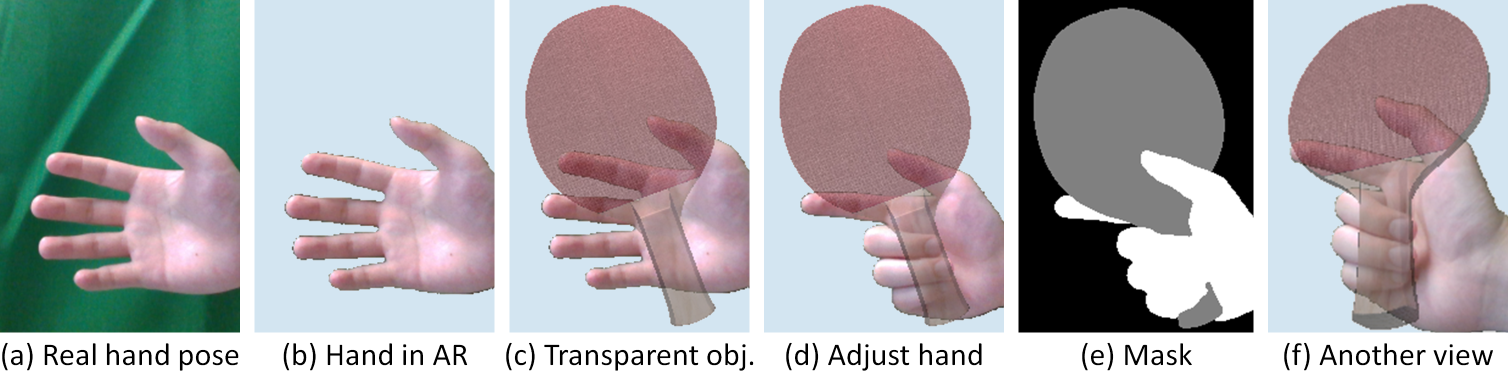}
	\caption{Real data preparation.
		(a) Captured hand in real physical space.
		(b) Rendered hand in virtual space.
		(c) Semi-transparent virtual object overlaid on hand.
		(d) Adjusted hand pose to grab the virtual object.
		(e) Manually-labeled occlusion mask (on hand).
		(f) After we grab-and-rotate the virtual object with our real hand.}
	\label{fig:real_data}
	\vspace*{-0.5mm}
\end{figure}


\section{Datasets}
\label{sec:data}


Our dataset contains image tuples of
(i) hand;
(ii) virtual object; and
(iii) associated occlusion mask (as ground truth), which marks the visible portions of hand and object.
%
To prepare compatible images (i)-(iii) in the same view, a naive approach is to manually label the visible regions in hand and virtual object images.
However, such a process is very tedious and the hand pose may not match the virtual object in the AR view.
Another approach is to use an RGBD sensor, but
the acquired depth is often noisy and may not match the rendered depth, so
we still need to manually edit the occlusion mask.

To address these challenges and obtain a large amount of data, 
we prepare a large synthetic data and a small real data, and design a two-stage procedure to train our neural network: {\em first, we use the synthetic data to pre-train the network, then we use the real data to fine-tune it\/}.
The key advantage behind is that we can efficiently obtain large synthetic data with automatically-generated occlusion masks without the burden of manual labeling, and the real data can further help to address the domain gap due to the synthetic renderings.

\textbf{Synthetic dataset.}
Figures~\ref{fig:synthetic_data} (a)-(f) illustrate how we prepare the synthetic data.
%
First, we use LeapMotion~\cite{leapmotion} to capture the 6-DoF skeletal joints of hand in physical space (a) and deform a rigged 3D hand model to fit these skeletal joints (b).
Then, we render the virtual object around the 3D hand model and display them together (c).
By adjusting the hand pose to appear to grab the virtual object, we can obtain a plausible hand-object occlusion in the virtual space (d).
We can then locate the foreground pixels of hand and virtual object (e) and automatically generate the occlusion mask (f).
By rendering the results in varying camera views and repeating the process for different target objects, we can efficiently produce a large volume of synthetic data with occlusion masks.

\textbf{Real dataset.}
%
Next, Figures~\ref{fig:real_data} (a)-(f) illustrate the procedure to prepare the real data.
First, we use a webcam to capture the hand (a) and a display to continuously show it in the virtual space (b).
We then render the target virtual object with transparency over the hand (c) to allow us to interactively adjust the hand pose until it appears to grab the object (d).
So, we can obtain a real image of the hand and a compatible rendered image of the virtual object in the same view, and label the visible portion of hand as the occlusion mask (e).
Further, to efficiently capture more data of different poses and views, we attach an ArUco marker~\cite{garrido2014automatic} below the hand to track its pose in the camera view.
Once the hand appears to grab the virtual object, we use the tracked pose information to move the virtual object with hand.
Therefore, we can capture more images of hand and object in different views (f).

\begin{figure}[!t]
	\centering
	\includegraphics[width=8.35cm]{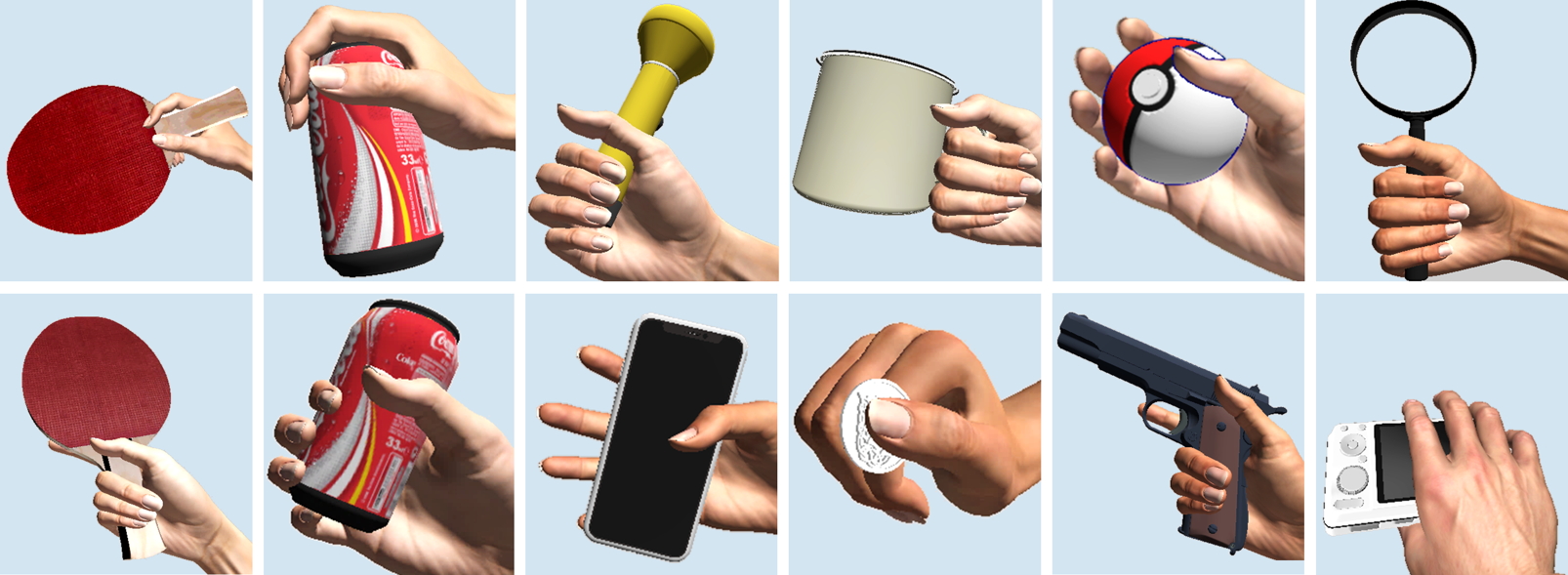}
	\caption{Virtual objects employed in preparing the datasets.}
	\label{fig:some_model}
\end{figure}


We employed ten virtual objects of various types when compiling the datasets; see Figure~\ref{fig:some_model}.
Note the different object grabbing poses, and some objects can have multiple (common) grabbing poses, e.g., see the ping-pong paddle in Figure~\ref{fig:some_model}.
Altogether, we collected 24,539 image tuples in the synthetic dataset and 1,232 image tuples in the real dataset.
We will release the two datasets upon the publication of this work.

\begin{figure*}[t]
	\centering
	\includegraphics[width=\textwidth*\real{0.95}]{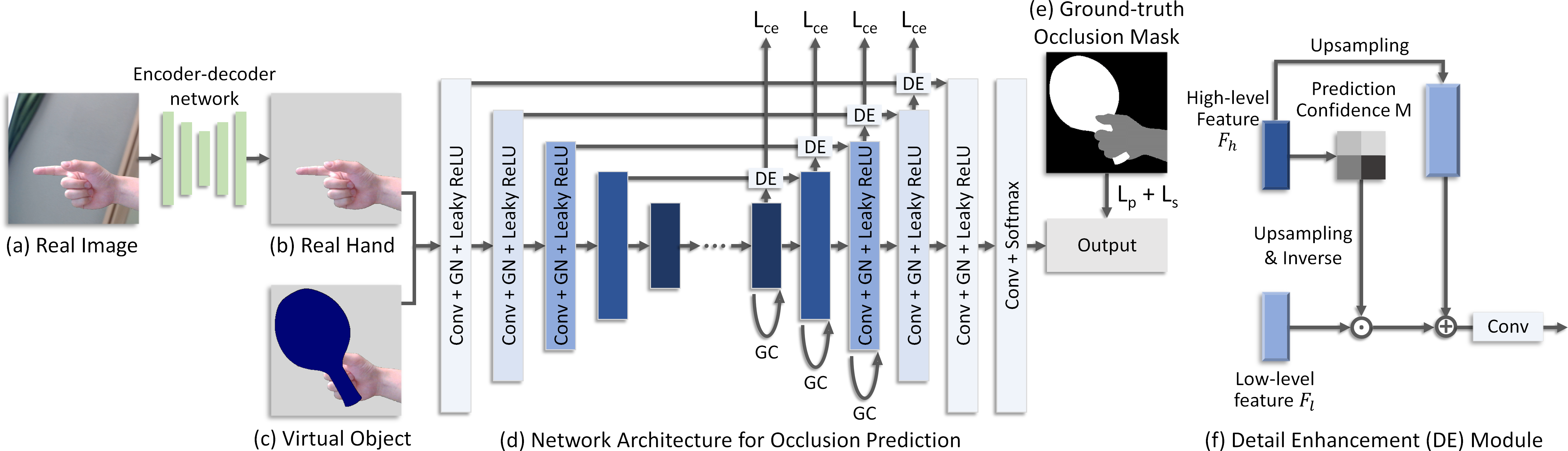}
	\caption{
		(a) Input real image;
		(b) hand extracted from real image;
		(c) virtual object;
		(d) our deep neural network architecture for occlusion prediction, in which Conv, GN, GC, and DE denote the convolutional operation, group normalization, global context module, and detail enhancement module, respectively; and
		(e) ground-truth occlusion mask from the prepared dataset for supervising the network training.
		Note that we employ the cross-entropy losses $L_{ce}$ as the supervisions for the intermediate layers in decoder, and the progressively-focusing cross-entropy loss $L_{f}$ and local smoothness loss $L_{s}$ to supervise the final network output.
		(f) The structure of the detail enhancement (DE) module.
	}
	\label{fig:network}
	\vspace*{-1.5mm}
\end{figure*}


\section{Method}

In this section, we present the architecture of our deep neural network and the implementation details and network losses we designed for improving the performance.

%

\subsection{Neural Network Architecture}
\label{sec:network}
Figures~\ref{fig:network} (a)-(e) show the whole neural network architecture.
%
We first extract the hand portion (b) in the input image (a) using an encoder-decoder network, and design another deep neural network architecture (d) to predict the occlusion mask.
The network (d) has a five-block encoder and a five-block decoder to extract convolutional features, where 
%
each block has a convolutional operation (Conv), a group normalization (GN), and a leaky ReLU operation.
Also, we introduce the global context modules (GC) in the first three blocks of the decoder to aggregate global information, 
%
and design the detail enhancement (DE) modules to harvest detail information in low-level features with skip connections.  
Further, we leverage another convolutional operation and softmax to generate the network output (occlusion mask), and use deep supervision to compute the loss and predict an occlusion mask per layer in the decoder. 
In the end, we take the occlusion mask predicted from the last layer as the final network output. 



\begin{figure}[!t]
	\centering
	\includegraphics[width=8.35cm]{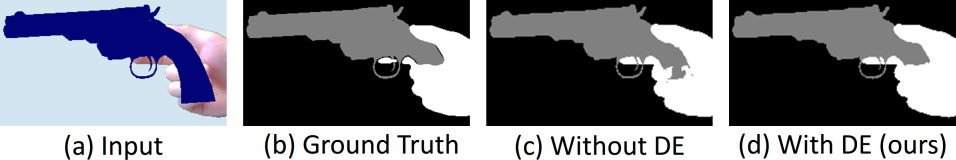}
	\caption{User grabs a virtual pistol (a) and the ground-truth occlusion relation is shown in (b). Compared to our network without DE (c), our network with DE (d) better preserves the details around the fingers.}
	\vspace{-1mm}
	\label{fig:de}
\end{figure}

\textbf{Global context module.}
Convolution operations tend to miss the global context, since they focus on local regions~\cite{wang2018non}.
However, to better infer the hand-object occlusions, we should treat the whole hand gesture as the global context and ignore the background.
Thus, we remove the background from the input hand image (see Figure~\ref{fig:network} (b)) using a simple encoder-decoder network, so the occlusion prediction network can focus on the hand (foreground). 
Further, we adopt the fast global context module (GC) in~\cite{cao2019gcnet} and compute the correlation among pixels in the feature map. Please see~\cite{cao2019gcnet} for details.
\textbf{Detail enhancement module.}
Low-level features in deep neural networks contain rich fine details.
Hence, to enhance the boundaries between hand and virtual objects, we formulate the detail enhancement (DE) module to harvest detail features $F_l$ at shallow layers of the network when the prediction confidence is low.
%
%
%
As shown in Figure~\ref{fig:network}(e), we extract the prediction confidence $M$ at deep layers $F_h$ via a convolutional layer and softmax, and obtain the enhanced low-level feature $F'_l$ by
\begin{equation}
F'_l \ = \ (2-\mathrm{Up}(M) \ \times \ 1.5) \ \times \ F_l \ ,
\end{equation}
where $\mathrm{Up}$ denotes the upsampling operation, which enlarges mask $M$ to the size of the feature maps at the corresponding shallow layer (see Figure~\ref{fig:network}(e)), and $(2-\mathrm{Up}(M) \ \times \ 1.5)$ aims to rescale the range of pixel value from $0.33\sim1$ (three classes in total) to $1.5\sim0.5$, which are then taken as weight to indicate the importance of the detailed features at shallow layers.
Lastly, we enlarge the high-level feature map $F_h$ at deep layer, add it to the refined low-level feature $F'_l$, and merge them through a $3\times3$ convolutional operation.
Note that our detail enhancement module introduces only the parameters of one convolutional operation, so the computational time is negligible. 
Figure~\ref{fig:de} shows the prediction results of the network with and without DE modules, where our method with DE better preserves the detailed structures in the prediction mask.

\subsection{Loss Functions}

\textbf{Progressively-focusing cross-entropy loss.}
\label{sec:loss}
To predict the hand-object occlusion, the key is to look at the overlap region between the virtual object and real hand.
Hence, we first formulate the focusing cross-entropy loss $L_f$ to emphasize more on the overlap regions in the network training:
\begin{equation}
\label{FCE}
L_f \ = \ -\sum^{H\times W}_{i}{(\alpha\Omega_i+\beta) \times y_i\log(p_i)} \ ,
\end{equation}
where
$\Omega$ is a mask that indicates the overlap region between the virtual object and real hand;
$\Omega_i$ equals one, if pixel $i$ is in the overlap region, and it equals zero, otherwise;
$\alpha$ and $\beta$ are for softening the hard-coded mask $\Omega$; 
$H\times W$ is the size of the target domain;
and $y_i$ and $p_i$ denote the ground truth and prediction result of pixel $i$, respectively.
%

\begin{figure}[!t]
	\centering
	\includegraphics[width=8.35cm]{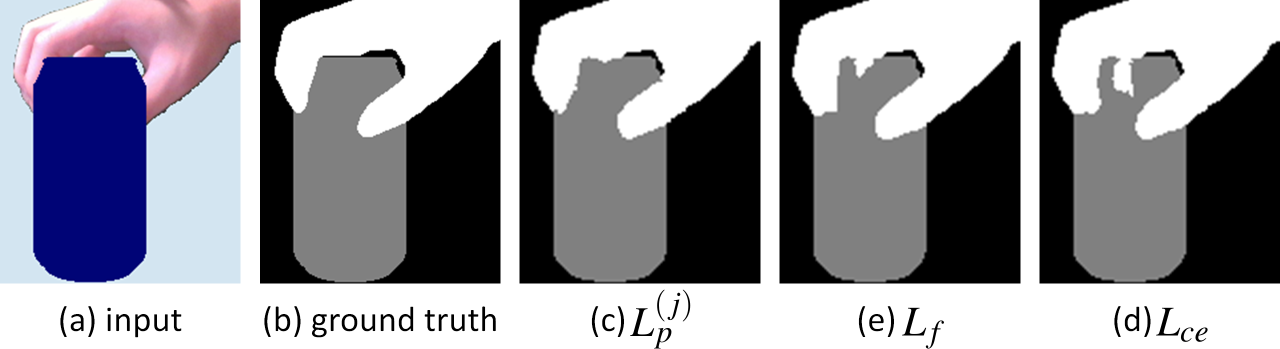}
	\caption{Visual comparison of 
	using (c) progressively-focusing cross-entropy loss $L_p^{(j)}$ in overall loss $L$,
	replacing $L_p^{(j)}$ in $L$ by
	(d) the focusing cross-entropy loss $L_f$.
	or
	(e) the standard cross-entropy loss $L_{ce}$.}
	\label{fig:pfl}
\end{figure}

However, the overlap region is usually a small part of the whole image. If the loss focuses too much on it, the network is hard to converge, due to the lack of global information.
Hence, we formulate a learning scheme to adaptively adjust the weight over time. The final progressively-focusing cross-entropy loss $L_p^{(j)}$ at the $j$-th training iteration is defined as
\begin{equation}
\label{PFCE}
L_p^{(j)} \ = \ -\sum^{H\times W}_{i}{\big[ \ (\frac{j}{N})^\tau [ \alpha \Omega_i - \beta ] + 1.0 \ \big] \times y_i\log(p_i)} \ ,
\end{equation}
where 
$N$ is the total number of training iterations;
$j$ is the current iteration ($j \in \{1, ..., N\}$), so
$j/N$ increases from zero to one following a polynomial function;
and parameters $\alpha$, $\beta$, and $\tau$ are empirically set as $0.4$, $0.2$, and $0.8$, respectively.
Hence, the network can gradually pay more attention to the regions of interest ($\Omega_i$), yet having a global perception of the whole image.
Figure~\ref{fig:pfl} shows the visual comparisons of using different loss terms, where the proposed progressively-focusing cross-entropy loss achieves the best performance.

\textbf{Local smoothness loss.}
We present a local smoothness loss to improve the smoothness of the boundaries in the occlusion predictions by minimizing the gradient orientation between the prediction and ground truth. 
First, we apply a soft-argmax function~\cite{chapelle2010gradient} on the network prediction $p$, and obtain a label map $b$, such that $b_i$ indicates the class of pixel $i$.
Here, we have three classes: background, object, and hand.
\begin{equation}
b_i \ = \ \sum^3_{k=1}{\frac{e^{\gamma p_{i,k}}}{e^{\gamma p_{i,1}}+e^{\gamma p_{i,2}}+e^{\gamma p_{i,3}}}k} \ ,
\end{equation}
where 
$p_{i,k}$ is the raw network output (before softmax) that indicates the probability of pixel $i$ being in class $k$; and
$\gamma$ is a parameter (which is set as 100 in practice) to enlarge the weight of the larger values.
%
Next, we obtain the gradient orientations $\vec{o}$ at label $b_i$ and at ground truth $y_i$ by  $3\times3$ Sobel operators, and compute the mean squared loss in the angular domain as the local smoothness loss:
\begin{equation}
L_s \ = \ \sum^{H\times W}_{i} ||\cos^{-1}\langle \vec{o}(b_i) \ , \ \vec{o}(y_i) \rangle||^2 \ .
\end{equation}

%

\textbf{Overall loss function.}
Finally, we combine the cross-entropy loss $L_{ce}$ defined on the prediction results produced from the four network layers (see $L_{ce}$'s in Figure~\ref{fig:network}), the progressively-focusing cross-entropy loss $L_p^{(j)}$, and the local smoothness loss $L_s$, to formulate the overall loss function
\begin{equation}
L \ = \ \sum^4_{s=1}{\alpha_1 L^s_{ce}} \ + \ \alpha_2 L_p^{(j)}  \ + \ \alpha_3 L_s \ ,
\end{equation}
where we empirically set parameters $\alpha_1$, $\alpha_2$, and $\alpha_3$ as $0.2$, $0.2$, and $\frac{1}{30}$, respectively.

\subsection{Training and Testing Strategies}

\vspace*{-0pt}
\textbf{Training parameters.}
First, we initialize the network parameters by random noise, which follows a zero-mean Gaussian distribution with a standard deviation of 0.1.
Then, we train the network by using $1:8$ as the ratio of real to synthetic data in each mini-batch. we empirically set the batch size as 16 and optimize the network by a stochastic gradient descent optimizer with a momentum value of 0.9 and a weight decay of $5\times10^{-4}$. 
The learning rate is adjusted by a polynomial strategy with a decay rate of 0.9. We terminate the whole training process after 50 epochs, and employ horizontal flip, random rotation, random sharpness, random brightness, and random contrast as the data augmentation.
Finally, we adopt the above strategies to first train the encoder-decoder network for real hand segmentation, then adopt the same strategies to jointly train this segmentation sub-network and the whole network for occlusion prediction. A median filter and a morphological close filter are then applied to the network output.

\vspace*{-0pt}
\textbf{Discussion.}
To meet the real-time processing requirement, the network model has to be simple and lightweight.
In the course of this work, we follow the design principle of ``fewer channels with more layers''~\cite{kim2016pvanet} to build the network, which consumes less processing time and produces higher accuracy.
%
On the other hand, in the early development of this work, we trained a network model individually for each virtual object in the dataset.
The results were, however, not satisfactory.
Interestingly, we later trained the network using the whole dataset of all virtual objects together.
The results improved, since more training data and object categorizes enhance the generalization capability of the deep neural network. 
\begin{figure}[!t]
\centering
\includegraphics[width=7.8cm]{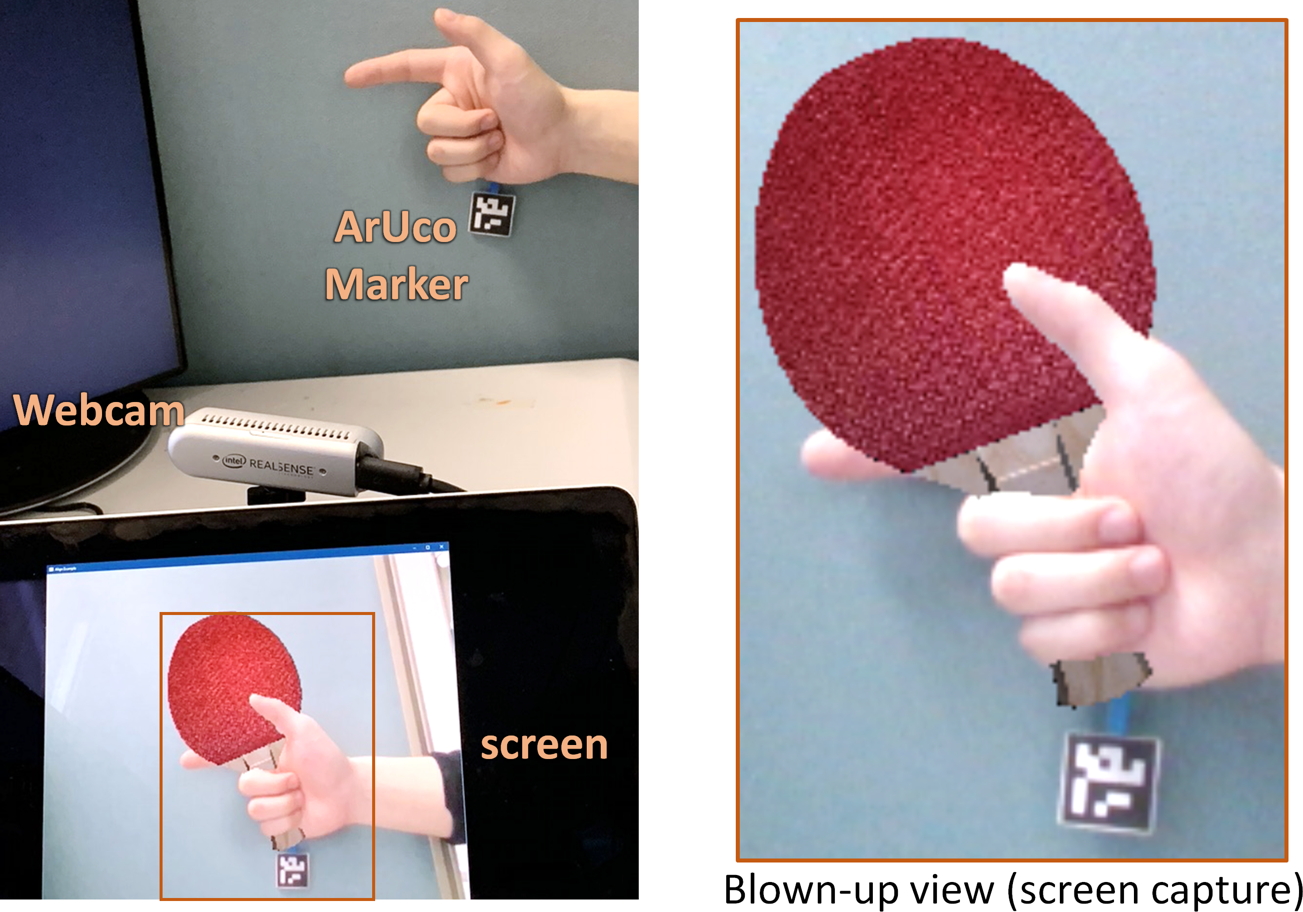}
\caption{Our prototyping AR System.}
\label{fig:proto_system}
\vspace{-2mm}
\end{figure}

\section{Prototyping AR System and Applications}
\label{sec:ARsystem}

With plausible occlusions between the virtual object and real hand, the virtual object can become grabbable and movable, as well as functional. We develop a prototyping AR system to explore the design space of functional virtual objects.
Figure~\ref{fig:proto_system} shows the prototyping AR system with
(i) a webcam that live captures the hand in front of the camera;
(ii) an ArUco marker attached below the hand for tracking; and
(iii) a screen that shows the composed AR view.
%
%
%
To track user's hand, initially we utilized several RGB-based 3D hand pose estimation networks~\cite{ge20193d,zimmermann2017learning}. However, these methods did not fit our task for several reasons, including not being able to meet the real-time requirement and weak robustness for smooth interactions. Thus, we use an ArUco marker in our prototyping system to track the hand pose in front of the camera. We will explore other methods for hand pose tracking in the future.

In detail, we render the virtual object, as if it is at a distance of $\sim$40cm from the camera.
When the user's hand enters the camera view, our system will start to use the neural network to determine the occlusion mask and take it to compose the hand and virtual object in the view.
Then, when the user feels that the object has been grabbed, he can use the other hand (the non-dominant one) to press a button to tell the system that the object has been grabbed.
We leave it as a future work to automatically tell the moment at which the real hand grabs the virtual object in the AR view.
After the grab, the pose of the virtual object will be locked with the hand, so we can use bare hand to directly move it and further perform various interesting interaction scenarios, as presented below:

\vspace*{-0.16mm}
\textbf{Scenario \#1: Virtual but \emph{clickable} lightsaber.}
The first scenario features the virtual lightsaber shown in Figure~\ref{fig:scenarios} (a).
It is a virtual 3D object in the AR view with a clickable button.
After one grabs the object with bare hand and presses the (red) button on it, the light blade can show up over the lightsaber. This is an unseen virtual object, not in the training data.


\vspace*{-0.16mm}
\textbf{Scenario \#2: Virtual but \emph{zoom-able} loupe.}
Figure~\ref{fig:scenarios} (b) shows a virtual magnifying glass (loupe) that allows us to magnify the background in the glass.
When one grabs this virtual loupe, he/she can switch between normal and magnifying modes by touching the glass surface with his/her thumb.


\vspace*{-0.16mm}
\textbf{Scenario \#3: Virtual but \emph{shoot-able} camera.}
Figure~\ref{fig:scenarios} (c) shows our virtual camera, with which one can grab, aim at a target, and press the shuttle to take pictures; here, a flash can also be triggered.
Besides, we can use this camera to reveal hidden contents on screen; see again Figure~\ref{fig:scenarios} (c),~\eg, the computer screen is grey but colors are shown on the camera.



\vspace*{-0.16mm}
\textbf{Scenario \#4: Virtual but \emph{scrollable} phone.}
The fourth scenario is a virtual phone in AR; see Figure~\ref{fig:scenarios} (d).
After one grabs the phone in AR, our system can present the hand (real) and phone (virtual) with natural partial occlusions.
Further, when one uses his/her thumb to scroll up/down, the virtual phone can respond to the action like a real one on our hand.
We develop a simple UI on the virtual phone, with which we can fade between real and virtual backgrounds using the red slider and change the camera filters using the blue slider.
%
With see-through AR glasses in the future, this interaction scenario suggests that one may manipulate a smartphone in the AR view, even without physically holding a real phone.


\vspace*{-0.16mm}
\textbf{Scenario \#5: Virtual but \emph{manipulatable} walkie-talkie.}
The fifth scenario features a virtual walkie-talkie (see Figure~\ref{fig:scenarios} (e)), with which one can press the red button with thumb to turn it on and scroll on its side with his/her index finger to switch channels. This virtual object is not in our training data.


\vspace*{-0.16mm}
\textbf{Scenario \#6: Virtual but \emph{rotatable} knob.}
The sixth scenario features a virtual but grabbable and rotatable knob (see Figure~\ref{fig:scenarios} (f)).
After one grabs it in the AR view, he/she can rotate it to the left or right.
Also, one may release the knob, grab it again, and make a larger turn until he/she opens the safe.
This virtual object is also not in our training data.



The above scenarios show that empowered by our prototyping system, we can design AR objects that are not only grabbable but also functional (clickable, scrollable, etc.) and greatly improve the user experience when interacting with virtual objects in AR, beyond what the existing AR applications can offer.
{\em Please refer to the supplementary video for demonstrations.\/}

%

\begin{figure*}[!t]
	\centering
	\includegraphics[width=0.99\textwidth]{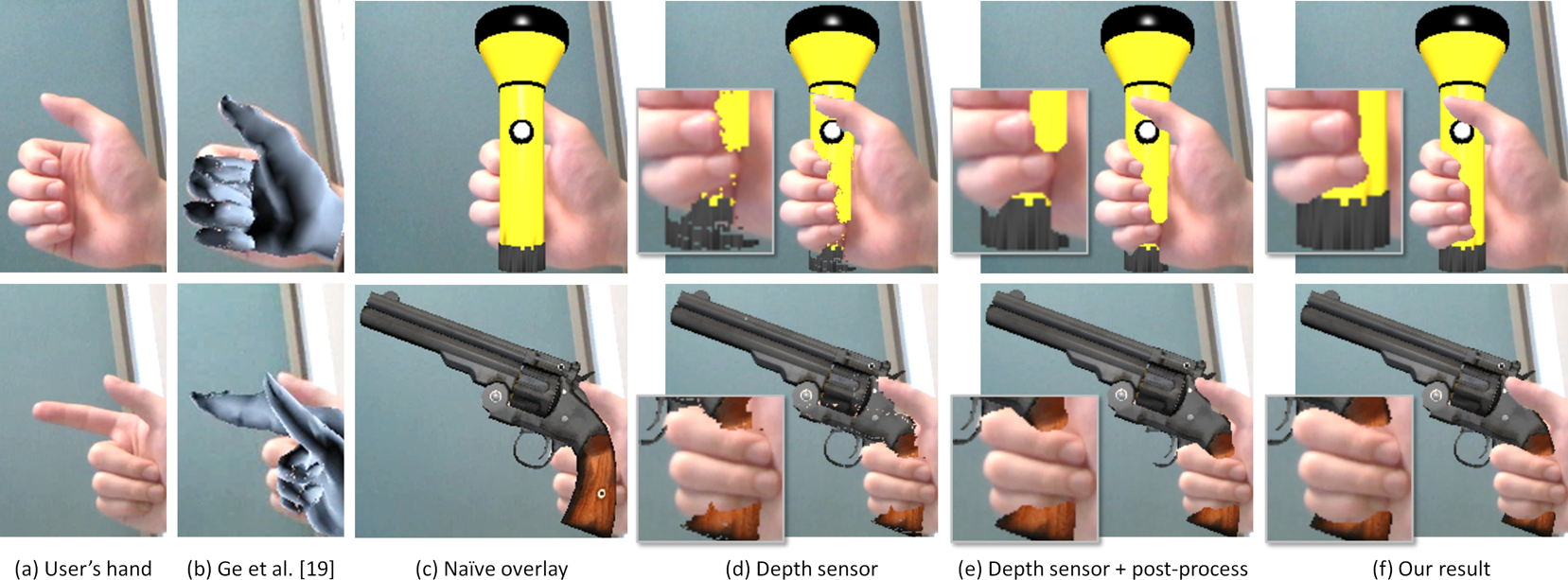}
	\caption{Visual comparisons between different methods on typical seen (above) and unseen (below) test data.}
	\label{fig:comparison_all}
\end{figure*}

\section{Evaluations}

\label{section:experiment}
This section presents a series of experiments to evaluate GrabAR, including time performance, qualitative and quantitative comparison with other approaches, and two user studies to explore the user interaction experience with occlusions.



\subsection{Time Performance}
GrabAR is built on a desktop computer with a 3.7GHz CPU and a single NVidia Titan Xp GPU.
Also, it makes use of the Logitech C920 camera to capture RGB images of the real world. 
%
The system runs at $32.26$ frames per second for $320\times320$ images, including the image capture.
Table~\ref{time_consumption} shows the detailed time performance for each computation stage.

\begin{table}[!t]
	\vspace*{-1mm}
	\caption{Time performance for each computation stage.}
	\label{time_consumption}
	\centering{
		\resizebox{7.0cm}{!}{
			\begin{tabular}{C{6.0cm}C{1.7cm}}
				\hline
				Stages             	& Time   \\ \hline
				(i) Inputs (capture hands \& render)	& 10ms  \\
				(ii) Network for real-hand extraction 	& 7ms  \\
				(iii) Network for occlusion prediction      	& 8ms  \\
				(iv) Image composition              	& 6ms  \\ \hline
				Total							& 31ms  \\
			\end{tabular}
	}}
	\vspace*{-2mm}
\end{table}

\subsection{Visual Comparison on Seen and Unseen Objects}
Existing methods~\cite{battisti2018seamless,feng2018resolving} mostly adopt depth sensors to estimate the occlusion mask between the real hand and virtual objects. 
Here, we use the Intel RealSense D435 depth camera to set up an AR system, in which we render the virtual objects over the real RGB background from a color camera, calibrate the color camera with the virtual camera, then use the depth information of the virtual objects (directly from rendering the object) and real hand (solely from the depth sensor) to determine their occlusion relationship.
%

Figure~\ref{fig:comparison_all} shows some of the visual comparison results.
The hand models from Ge~\etal~\cite{ge20193d} (see column (b)) do not align precisely with the hand, the ``na\"ive overlay'' approach (see column (c)) fails to give the perception that the real hand and virtual objects co-exist in the same space, whereas the ``depth sensor'' approach (see column (d)) introduces artifacts on the hand-object boundaries. 
Note, for better comparison, we applied the same post-processing step (see the ``Method'' section) to the results produced by the ``depth sensor'' approach (see column (e)), since the raw results contain heavy noise. 
In contrast, our method is able to generate more realistic composition of hand and virtual objects (see column (f)).
Further, note that the result in Figure~\ref{fig:comparison_all} (bottom) is on an unseen virtual object, which is not in our synthetic and real datasets; yet, our method can still produce plausible occlusions.
This result demonstrates the generalization capability of our method (see also the experiment in next subsection).
Since we trained GrabAR-Net using all the virtual objects in the training data together, it should be able to better learn the hand grabbing gestures rather than simply remembering the virtual objects.
Furthermore, our system is flexible to set up, comparing with those that rely on depth sensors. 
%
Note that more visual comparison results can be found in the supplementary material.

\subsection{Quantitative Comparison on Seen and Unseen Objects}
\label{sec:quantitative}

\vspace*{0.5mm}
\textbf{Evaluation metric.} \
Since the challenges of handling occlusions in AR happen near the hand-object boundaries, we thus design a new evaluation metric, namely ODSC, specifically to look at the overlap regions between hand and virtual object, while ignoring remaining regions in the image space.
To do so, we first extract overlap region ($\Omega$) between the real hand and virtual object, then compute the dice score coefficient (DSC) between the predicted occlusion mask ($P$) and ground truth image ($G$) within $\Omega$. 
In detail, we define ODSC as
\begin{equation}
ODSC = \frac{2 | P \cap G |_{\Omega}}{ |P|_\Omega + |G|_\Omega} \ ,
\end{equation}
where subscript $\Omega$ counts only pixels in $\Omega = H \cap O$, while $H$ and $O$ denote the hand and virtual object regions, respectively.

\vspace*{0.5mm}
\textbf{Evaluation results.} \
We recruited ten volunteers to help us annotate 180 pairs of real hand and virtual object images as the ground-truth.
This additional evaluation dataset covers ten different virtual objects: two have two different grabbing poses and three are unseen.
Then, we quantitatively compare the results produced by GrabAR-Net (GN) with results produced from the depth-sensor-based approach (DS), depth-sensor-based approach with post-processing (DSP) and Ge~\etal~\cite{ge20193d}.
Table~\ref{table:quantitative} reports the ODSC values for different seen and unseen virtual objects, showing that GrabAR-Net {\em consistently\/} produces far better results than the other approaches by a large margin.
%
We also conducted an ablation study to evaluate the effectiveness of the major components in our method. Please refer to the supplementary material for more details.

\begin{table}[!t]
	\caption{Quantitative comparison between different methods (DS: depth-sensor-based approach; DSP: depth-sensor-based approach with post-processing; and GN: GrabAR-Net) in terms of the ODSC metric.
	Unseen virtual objects are marked with ``*'' and virtual objects with different grabbing poses are marked with ``$^{\circ}$''.}
	\vspace{1.5mm}
	\label{table:quantitative}
	\centering{
		\resizebox{7.8cm}{!}{
			\begin{tabular}{C{1.35cm}|C{1.5cm}C{1.5cm}C{2.0cm}C{1.3cm}}
				\hline
				Object     & DS      & DSP    & Ge~\etal~\cite{ge20193d} & GN (our)           \\ \hline
				can$^{\circ}$ 	   & 63.35\% & 59.64\% & 73.56\% & \textbf{90.44\%}   \\
				cup        & 84.74\% & 63.24\% & 83.17\% & \textbf{87.14\%}   \\
				flashlight & 76.09\% & 72.72\% & 84.97\% & \textbf{95.30\%}   \\
				gun        & 80.78\% & 79.57\% & 84.67\% & \textbf{96.42\%}   \\
				loupe      & 91.13\% & 80.61\% & 86.25\% & \textbf{94.41\%}   \\
				paddle$^{\circ}$    & 84.78\% & 80.68\% & 70.56\% & \textbf{95.02\%}   \\
				phone      & 85.16\% & 84.48\% & 86.61\% & \textbf{95.78\%}   \\ \hline
				knob*      & 76.89\% & 75.72\% & 66.64\% & \textbf{81.79\%}   \\
				lightsaber*& 84.84\% & 83.23\% & 78.98\% & \textbf{94.46\%}   \\
				pistol*    & 93.91\% & 89.51\% & 84.90\% & \textbf{96.73\%}   \\ \hline
				average    & 81.57\% & 75.81\% & 78.69\% & \textbf{92.76\%}
	\end{tabular}}}
	\vspace{-1mm}
\end{table}

%
%



\begin{figure}[!t]
	\centering
	\includegraphics[width=8.35cm]{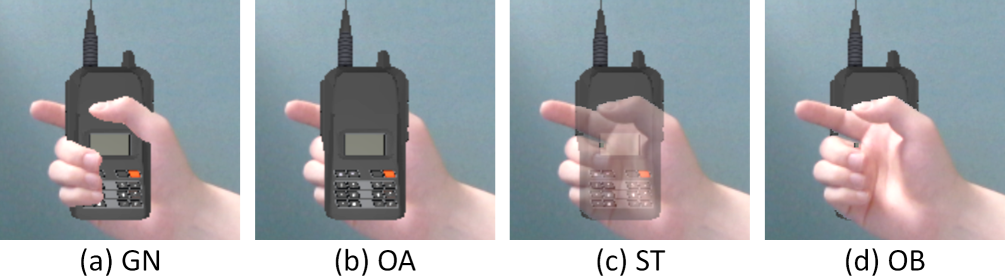}
	\caption{Different ways of handling the occlusion between virtual object and real hand:
	(a) our GrabAR-Net (GN);
	(b) render virtual {\em object above\/} user's hand (OA);
	(c) make the hand-object overlap region semi-transparent (ST); and
	(d) render virtual {\em object behind\/} user's hand (OB).}
	\label{fig:user_study}
	\vspace*{-2mm}
\end{figure} 

\subsection{Study 1: The Effect of Occlusion}
First, we explore how occlusions affect user interaction experience. 
We compare GrabAR-Net (GN) with (i) rendering the virtual \emph{object above} user's hand (OA) and (ii) making the hand-object overlap region \emph{semi-transparent} (ST); see Figure~\ref{fig:user_study}.
%
Before this study, we recruited three participants in a pilot study, in which we considered also rendering the virtual \emph{object behind} user's hand (OB).
All participants reported that their hands occluded most of the virtual object in OB and hindered their interactions, so we ignored OB in Study 1. 

\vspace*{0.5mm}
\textbf{Participants.}
We recruited nine participants in campus: four females and five males, aged between 22 and 28.
Among them, eight had experience with AR, and all are right-handed.

\textbf{Tasks.} \
\noindent
Task 1 is on \emph{rotating the virtual knob} shown in Figure~\ref{fig:scenarios}.
In each trial, the participant presses a button to initiate the trial, then adjusts his/her hand to grab the virtual knob. 
When the participant feels like successfully grabbing the knob in the AR view, he/she can press another button and start rotating the knob for $+90$, $-90$, and $+70$ degrees to open the safe.
%
We record the time taken by each participant to adjust hand for grabbing and also the time to complete the task.

Task 2 is on \emph{operating the virtual walkie-talkie} shown in Figure~\ref{fig:scenarios}.
In each trial, the participant opens the walkie-talkie and uses his/her index finger to freely switch channels with no time limit. 
To open the walkie-talkie, the participant can use his/her thumb to touch the red button via ST/GN, or press a button on keyboard via OA.
Like \emph{(i)}, we record the object grabbing time.
Note that, in both tasks, we counterbalance the order of the compared methods across the participants.

\vspace*{0.5mm}
\textbf{Procedure.} \
When a participant came to our lab, we first introduced the system to him/her and started a ten-minute tutorial for him/her to try out these tasks for all methods.
Then, the participant performed the tasks.
After that, each participant was asked to fill a questionnaire with two Likert-scale questions (1: disagree and 5: agree) on GN, ST, and OA:
(Q1) the AR object is plausible and realistic (``AR object'' for short); and
(Q2) the interaction experience is intuitive and immersive (``Interaction'' for short).
Lastly, we interviewed the participant to obtain his/her comments.

\begin{figure}[!t]
	\centering
	\includegraphics[width=8.35cm]{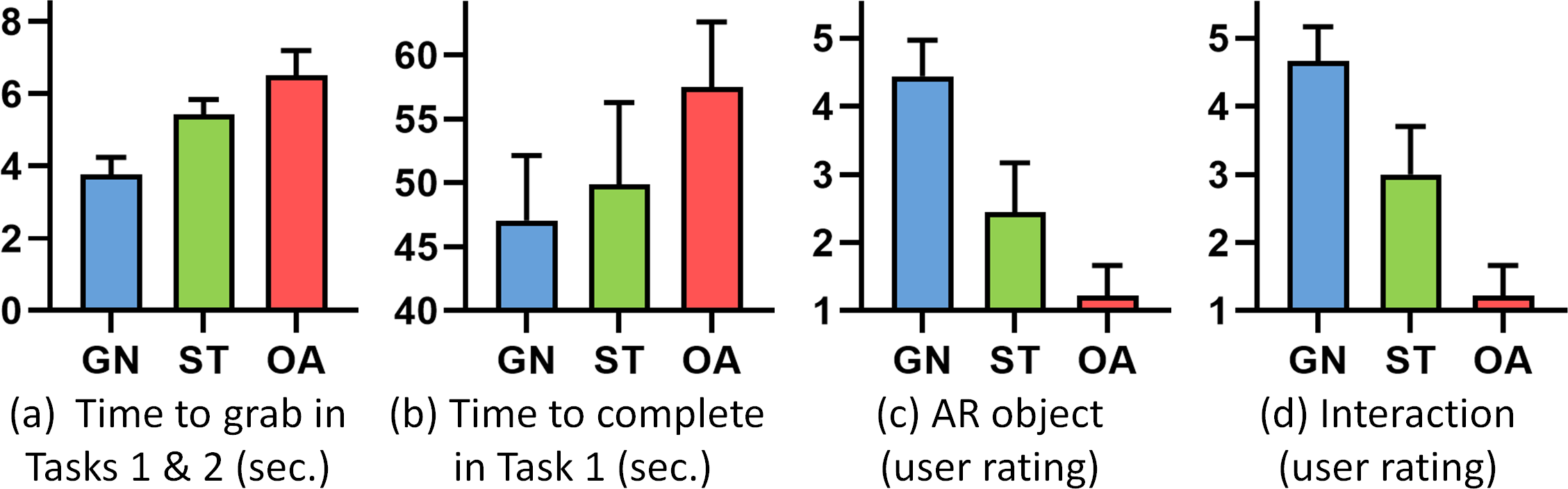}
	\caption{Study 1 results.
	Comparing our GrabAR-Net (GN) with rendering virtual object above user's hand (OA) and making the hand-object overlap region semi-transparent (ST). Note that the boxes show the mean values (M) and error bars show the standard deviations (SD), whereas the user ratings range 1 (worst) to 5 (best).}
	\label{fig:user_study_result}
	\vspace*{-1.5mm}
\end{figure}

\vspace*{0.5mm}
\textbf{Quantitative comparison results.} \
Figure~\ref{fig:user_study_result} summarizes the results, in which we further evaluate the significance using paired t-tests on the first two time-related objective quantities and the Wilcoxon signed rank tests on the user ratings:

\vspace*{-0.17mm}
\emph{(i) Time to grab (sec.).} \
Time to grab the virtual objects under GN (M=3.766, SD=0.455) is significantly shorter than those of ST (M=5.419, SD=0.398, t-test: t(8)=6.337, p=0.0002) and OA (M=6.514, SD=0.641, t-test: t(8)=8.059, p<0.0001), where
$M$ and $SD$ are mean and standard deviation, respectively. 

\vspace*{-0.17mm}
\emph{(ii) Time to complete Task 1 (sec.).} \
Completion time under OA (M=57.52, SD=4.776) is significantly longer than those under GN (M=47.08, SD=4.787, t-test: t(8)=5.107, p=0.0009) and ST (M=49.87, SD=6.049, t-test: t(8)=3.690, p=0.0061), whereas GN and ST have no significant differences.


\vspace*{-0.17mm}
\emph{(iii) AR object (Q1).} \
User ratings under GN (M=4.8, SD=0.416) are significantly higher than those under OA (M=1.2, SD=0.416, Z=-2.739, p=0.006) and ST (M=2.4, SD=0.685, Z=-2.716, p=0.007).
The result implies that the participants found GN producing better AR composition.

\vspace*{-0.17mm}
\emph{(iv) Interaction (Q2).} \
User ratings under GN (M=4.6, SD=0.497) are significantly higher than those under OA (M=1.2, SD=0.416, Z=-2.724, p=0.006) and ST (M=3.0, SD=0.667, Z=-2.714, p=0.007).
The result implies that participants preferred the interaction experience under GN.

\vspace*{0.5mm}
\textbf{Participant comments.} \
Almost all participants stated that OA is not ideal. Speaking of ST, P3 said that ``ST made her feel that she was moving fingers behind the virtual object. P7 explained why it costed him more time to grab the virtual object under OA and ST: ``Without occlusion, I was not sure if my grabbing pose was correct, so I hesitated for a while before grabbing.'' The participants highly praised GN: Plausible occlusion ``made me feel that I was actually grabbing the virtual object'' (P3) and ``brought fancy AR scenes'' (P1 \& P4).

\subsection{Study 2: The Quality of Occlusion}

%
Next, we compare GN with depth-sensor-based methods on the quality of resolving hand-object occlusion.
%
Since Ge~\etal~\cite{ge20193d} is not real-time (10.75 FPS), we excluded it in Study 2. 
Also, in a pilot study, we found that the participants always preferred the depth-sensor-based method with post-processing (DSP), which has less noise, than the vanilla depth-sensor-based method, 
so we only compared GN with DSP here.
 
%
\vspace*{0.5mm}
\textbf{Participants.} \
We recruited ten participants in campus: four females and six males, aged between 22 and 28.
Among them, eight had experience with AR, and all are right-handed.


\vspace*{0.5mm}
\textbf{Tasks.} \
Task 1 is on grabbing virtual objects. 
Each participant was asked to grab six virtual objects of various kinds and poses (paddle with handshake and penhold grips, can, pistol, flashlight, and lightsaber).
The virtual objects were rendered using GN or DSP in different trials (two trials per permutation).
Overall, we have 240 trials: ten participants $\times$ six objects $\times$ two modes (GN/DSP) $\times$ two blocks of trials.
In each trial, the participant presses a button on the keyboard to initiate a trial.
Then, the system renders the next virtual object in the AR view using GN or DSP for the participant to stretch hand and make a hand grabbing pose to fetch the object.
Immediately after the participant feels that the object has been grabbed, he/she can use the other hand to press a button.
Our system then records the time taken by the participant in the trial.

Task 2 is on scrolling on the virtual phone with the same set up as Task 1, but focuses on letting the participants try and experience hand interactions with AR objects under GN and DSP.
With our system, the participant can grab the virtual phone, scroll on it like a real one, and look at different parts of the image on the phone screen; see Figure~\ref{fig:scenarios}.
Further, the participant can find a button and click on it.
In Task 2, we do not impose any time limit but require them to interact with GN and DSP for roughly the same amount of time.
Also, we count the number of times that their hand penetrates the virtual objects for GN and for DSP.
In both tasks, we counterbalance the order of GN and DSP across participants.

\vspace*{0.5mm}
\textbf{Procedure.} \
When a participant came to our lab, we first introduced the AR system and gave a tutorial session to him/her on how to grab a virtual object with na\"ive overlay, GN, and DSP.
Particularly, when we showed the na\"ive overlay mode to the participants, we asked them to point out the occlusion issue to ensure that they understood the occlusion relationship between the real hand and virtual objects.
After that, each participant performed Tasks 1 and 2 as presented above.

Then, each participant was asked to fill a questionnaire with three Likert-scale questions (1: disagree and 5: agree) on GN and DSP:
(Q1) the presented occlusion is handled correctly (``Correctness'' for short);
(Q2) the image quality is good, i.e., noise-free without flickering (``Image Quality'' for short); and
(Q3) the hand-object interaction is plausible and immersive (``Interaction'' for short).
Lastly, we further interviewed the participant to obtain their comments.

\begin{figure}[!t]
	\centering
	\includegraphics[width=8.35cm]{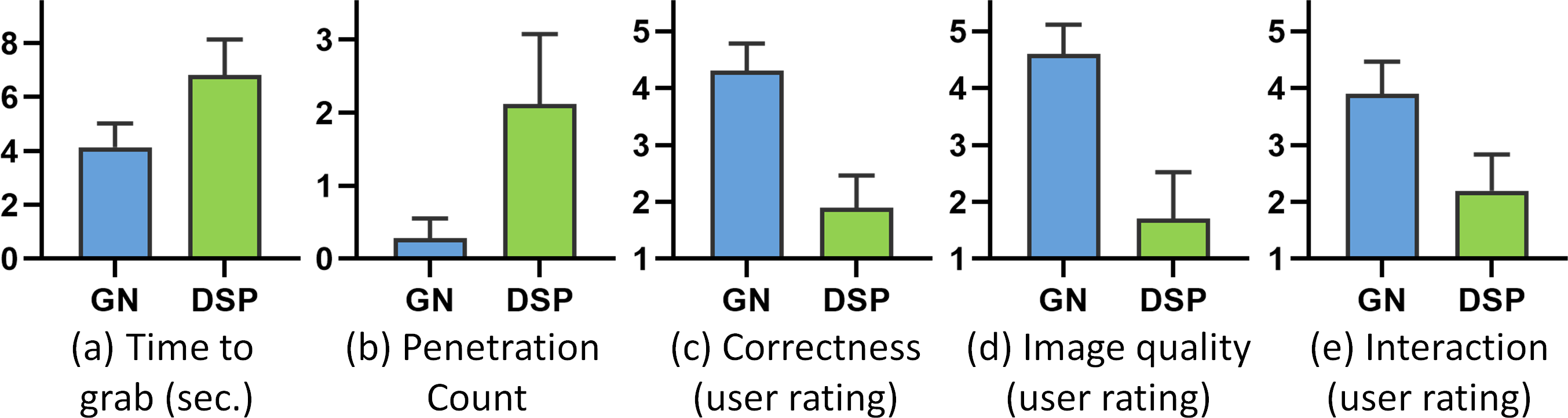}
	\caption{Study 2 results.
	Comparing our GrabAR-Net (GN) with the depth-sensor-based method with post-processing (DSP).
	Note that the boxes show the mean values (M), error bars show the standard deviations (SD), and user ratings range 1 (worst) to 5 (best).}
	\label{fig:user_study2_result}
\end{figure} 

\vspace*{0.5mm}
\textbf{Quantitative comparison results.} \
Figures~\ref{fig:user_study2_result} summarizes the results.
Again, we evaluate the significance using paired t-tests on the first two time-related objective quantities and the Wilcoxon signed rank tests on the user ratings.

\vspace*{-0.25mm}
\emph{(i) Time to grab the object} (in sec.) in Task 1 under GN is (M=4.135, SD=0.894), which is significantly shorter than those of DSP (M=6.820, SD=1.314, t-test: t(9)=13, p<0.0001).

\vspace*{-0.25mm}
\emph{(ii) Penetration count} under GN is (M = 0.28, SD = 0.270), which is significantly lower than those of DSP (M=2.12, SD=0.953, t-test: t(9)=6.241, p=0.0002).
The result shows that GN can better avoid object-hand penetration in the AR view, i.e., less interference during the hand-object interactions.


\vspace*{-0.25mm}
\emph{(iii) Correctness (Q1)} ratings under GN (M=4.3, SD=0.483) are significantly higher than those under DSP (M=1.9, SD=0.567, Z=-2.848, p=0.004), implying that the participants found the occlusions under GN to be more plausible.

\vspace*{-0.25mm}
\emph{(iv) Image Quality (Q2)} ratings under GN (M=4.6, SD=0.516) are significantly higher than those under DSP (M=1.7, SD=0.8233, Z=-2.836, p=0.005).
The result implies that GN produced better image quality as observed by the participants.

\vspace*{-0.25mm}
\emph{(v) Interaction (Q3)} ratings under GN (M=3.9, SD=0.568) are significantly higher than those under DSP (M=2.2, SD=0.633 Z=-2.701, p=0.007).
The result implies that participants had better interaction experience with GN.

\vspace*{0.5mm}
\textbf{Participant comments.} \
In Task 1, all participants found na\"ive overlay not ideal for supporting virtual object grabbing in AR and preferred GN, because GN offers sharper and more accurate results at edges, while the results of DSP have more noise and flickering.
%
As for Task 2, we observed that the participants often needed more time to adjust their hand poses to grab the virtual phone under DSP.
P2 gave a representative comment: ``It was hard for me to sense the distance between the camera and virtual object, so I often reached out my hand to the front or to the back of the virtual object instead of directly grabbing it between my fingers under DSP. But this situation happened less under GN.''
When scrolling on the virtual phone, seven participants pointed out that object-hand penetrations happened too often under DSP, due to unstable hand motions.
It is ``caused by the unintentional scrolling operations on the virtual phone (P1 \& P3).'' and ``interfered me and made me feel unrealistic (P6).''
Object-hand penetration also happens under GN, due to mis-predictions by GrabAR-Net, but ``it occurred at an acceptable frequency (P9).''

However, GN is not perfect all the time.
Two participants reported that attaching the ArUco marker to their hands was bothering, which is a limitation of our existing prototyping system.
Three participants mentioned that although the system was occlusion-aware, they still felt unrealistic due to the lack of tactile feedback, which is a common drawback in AR interactions.
However, P5 said that ``I preferred this comparing to carrying an extra physical device for tactile feedback.''

\section{Discussion and Future Work}

This paper presents GrabAR, a new approach
that makes use of an embedded neural network to directly predict real-and-virtual occlusion in AR views, while bypassing inaccurate depth estimation and inference.
Using our approach, we can enable users to grab and manipulate various virtual objects with accurate occlusion, and increase the sense of immersion for hand interactions with virtual objects.
Our key idea is to learn to determine the occlusion from hand grabbing poses that describe natural virtual object grabs with user's hand in physical space.
Particularly, we design a GrabAR-Net to predict the occlusion mask of the virtual object and real hand by formulating the global context module and detail enhancement module to aggregate the global and local information simultaneously and presenting two novel loss functions to provide better supervision for the network training.
Further, we compile synthetic and real datasets and use them to train the neural network to leverage the automatically-generated labels in synthetic data and reduce the burden of manually labeling the real data.
A prototyping AR system is presented with various interaction scenarios, showing that virtual objects in AR can be grabbable, as well as interactable.
We compared GrabAR with other methods through various experiments, and show that GrabAR provides significantly more plausible results than other existing methods, both visually and quantitatively.

As the first work, we are aware of, that explores the deep neural network to compose virtual objects with real hands, there are still several problems that we plan to work on in the future.
(i) Single hand interactions.
Currently, our system supports interaction using a single hand only.
We plan to extend it to allow bimanual interactions by including more training samples.
(ii) ArUco marker.
Our system relies on an ArUco marker to detect the six-DoF poses of user's hand, which is cumbersome to set up and limits the motion range of user's hand.
In the future, we will explore the use of only a single RGB image by training a deep neural network to learn the six-DoF hand poses automatically. 
(iii) Capturing user's intention.
Without depth reference, it is hard to capture user's intention of grabbing or releasing the virtual object. In our prototype system we require users to press a button to indicate the moment of grabbing and releasing. It somehow breaks the sense of immersion. 
In the future, we plan to explore methods,~\eg, time-series analysis, to simplify the interactions.
Moreover, we consider further improving our method by taking the temporal information of video into the network for smoothing the results.

%

\bibliographystyle{sigchireference}
\bibliography{sample}


\begin{thebibliography}{00}


\ifx \showCODEN    \undefined \def \showCODEN     #1{\unskip}     \fi
\ifx \showDOI      \undefined \def \showDOI       #1{{\tt DOI:}\penalty0{#1}\ }
  \fi
\ifx \showISBNx    \undefined \def \showISBNx     #1{\unskip}     \fi
\ifx \showISBNxiii \undefined \def \showISBNxiii  #1{\unskip}     \fi
\ifx \showISSN     \undefined \def \showISSN      #1{\unskip}     \fi
\ifx \showLCCN     \undefined \def \showLCCN      #1{\unskip}     \fi
\ifx \shownote     \undefined \def \shownote      #1{#1}          \fi
\ifx \showarticletitle \undefined \def \showarticletitle #1{#1}   \fi
\ifx \showURL      \undefined \def \showURL       #1{#1}          \fi

\bibitem{anderson2003role}
{Barton~L Anderson}. 2003.
\newblock \showarticletitle{The role of occlusion in the perception of depth,
  lightness, and opacity.}
\newblock {\em Psychological Review\/} {110}, 4 (2003), 785.
\newblock


\bibitem{baek2019pushing}
{Seungryul Baek}, {Kwang~In Kim}, {and} {Tae-Kyun Kim}. 2019.
\newblock \showarticletitle{Pushing the Envelope for {RGB}-based Dense {3D}
  Hand Pose Estimation via Neural Rendering}. In {\em CVPR}. 1067--1076.
\newblock


\bibitem{battisti2018seamless}
{Caterina Battisti}, {Stefano Messelodi}, {and} {Fabio Poiesi}. 2018.
\newblock \showarticletitle{Seamless Bare-Hand Interaction in Mixed Reality}.
  In {\em ISMAR}. 198--203.
\newblock


\bibitem{benko2007balloon}
{Hrvoje Benko} {and} {Steven Feiner}. 2007.
\newblock \showarticletitle{Balloon selection: A multi-finger technique for
  accurate low-fatigue {3D} selection}. In {\em 2007 IEEE Symposium on {3D}
  User Interfaces}.
\newblock


\bibitem{boukhayma20193d}
{Adnane Boukhayma}, {Rodrigo~de Bem}, {and} {Philip~HS Torr}. 2019.
\newblock \showarticletitle{{3D} hand shape and pose from images in the wild}.
  In {\em CVPR}. 10843--10852.
\newblock


\bibitem{cai2018weakly}
{Yujun Cai}, {Liuhao Ge}, {Jianfei Cai}, {and} {Junsong Yuan}. 2018.
\newblock \showarticletitle{Weakly-supervised {3D} hand pose estimation from
  monocular {RGB} images}. In {\em ECCV}. 666--682.
\newblock


\bibitem{cao2019gcnet}
{Yue Cao}, {Jiarui Xu}, {Stephen Lin}, {Fangyun Wei}, {and} {Han Hu}. 2019.
\newblock \showarticletitle{{GCN}et: Non-Local Networks Meet Squeeze-Excitation
  Networks and Beyond}. In {\em ICCV Workshops}.
\newblock


\bibitem{caudell1992augmented}
{Thomas~P Caudell} {and} {David~W Mizell}. 1992.
\newblock \showarticletitle{Augmented reality: An application of heads-up
  display technology to manual manufacturing processes}. In {\em Hawaii
  Iternational Conference on System Sciences}, Vol.~2. 659--669.
\newblock


\bibitem{chapelle2010gradient}
{Olivier Chapelle} {and} {Mingrui Wu}. 2010.
\newblock \showarticletitle{Gradient descent optimization of smoothed
  information retrieval metrics}.
\newblock {\em Information Retrieval\/} {13}, 3 (2010), 216--235.
\newblock


\bibitem{choi2013ihand}
{Junyeong Choi}, {Jungsik Park}, {Hanhoon Park}, {and} {Jong-II Park}. 2013.
\newblock \showarticletitle{i{H}and: an interactive bare-hand-based augmented
  reality interface on commercial mobile phones}.
\newblock {\em Optical Engineering\/} {52}, 2 (2013), 1--11.
\newblock


\bibitem{chun2013real}
{Wendy~H Chun} {and} {Tobias H{\"o}llerer}. 2013.
\newblock \showarticletitle{Real-time hand interaction for augmented reality on
  mobile phones}. In {\em Proceedings of the 2013 international conference on
  Intelligent user interfaces}. 307--314.
\newblock


\bibitem{dorfmuller2001finger}
{Klaus Dorfmuller-Ulhaas} {and} {Dieter Schmalstieg}. 2001.
\newblock \showarticletitle{Finger tracking for interaction in augmented
  environments}. In {\em ISMAR}. 55--64.
\newblock


\bibitem{eigen2014depth}
{David Eigen}, {Christian Puhrsch}, {and} {Rob Fergus}. 2014.
\newblock \showarticletitle{Depth map prediction from a single image using a
  multi-scale deep network}. In {\em NIPS}. 2366--2374.
\newblock


\bibitem{engel2014lsd}
{Jakob Engel}, {Thomas Sch{\"o}ps}, {and} {Daniel Cremers}. 2014.
\newblock \showarticletitle{{LSD-SLAM}: Large-scale direct monocular {SLAM}}.
  In {\em ECCV}. 834--849.
\newblock


\bibitem{feng2018resolving}
{Qi Feng}, {Hubert~PH Shum}, {and} {Shigeo Morishima}. 2018.
\newblock \showarticletitle{Resolving occlusion for {3D} object manipulation
  with hands in mixed reality}. In {\em VRST}. 119.
\newblock


\bibitem{fu2018deep}
{Huan Fu}, {Mingming Gong}, {Chaohui Wang}, {Kayhan Batmanghelich}, {and}
  {Dacheng Tao}. 2018.
\newblock \showarticletitle{Deep ordinal regression network for monocular depth
  estimation}. In {\em CVPR}. 2002--2011.
\newblock


\bibitem{furukawa2017depth}
{Ryo Furukawa}, {Ryusuke Sagawa}, {and} {Hiroshi Kawasaki}. 2017.
\newblock \showarticletitle{Depth Estimation Using Structured Light
  Flow--Analysis of Projected Pattern Flow on an Object's Surface}. In {\em
  ICCV}. 4640--4648.
\newblock


\bibitem{garrido2014automatic}
{Sergio Garrido-Jurado}, {Rafael Mu{\~n}oz-Salinas}, {Francisco~Jos{\'e}
  Madrid-Cuevas}, {and} {Manuel~Jes{\'u}s Mar{\'\i}n-Jim{\'e}nez}. 2014.
\newblock \showarticletitle{Automatic generation and detection of highly
  reliable fiducial markers under occlusion}.
\newblock {\em Pattern Recognition\/} {47}, 6 (2014), 2280--2292.
\newblock


\bibitem{ge20193d}
{Liuhao Ge}, {Zhou Ren}, {Yuncheng Li}, {Zehao Xue}, {Yingying Wang}, {Jianfei
  Cai}, {and} {Junsong Yuan}. 2019.
\newblock \showarticletitle{{3D} Hand Shape and Pose Estimation from a Single
  {RGB} Image}. In {\em CVPR}. 10833--10842.
\newblock


\bibitem{hoiem2007recovering}
{Derek Hoiem}, {Andrew~N Stein}, {Alexei~A Efros}, {and} {Martial Hebert}.
  2007.
\newblock \showarticletitle{Recovering occlusion boundaries from a single
  image}. In {\em ICCV}. 1--8.
\newblock


\bibitem{holynski2018occlusion}
{Aleksander Holynski} {and} {Johannes Kopf}. 2018.
\newblock \showarticletitle{Fast Depth Densification for Occlusion-aware
  Augmented Reality}.
\newblock {\em ACM Transactions on Graphics (SIGGRAPH Asia)\/} {37}, 6 (Dec.
  2018), 194:1--194:11.
\newblock
\showISSN{0730-0301}


\bibitem{kim2016pvanet}
{Kye-Hyeon Kim}, {Sanghoon Hong}, {Byungseok Roh}, {Yeongjae Cheon}, {and}
  {Minje Park}. 2016.
\newblock \showarticletitle{{PVA}net: Deep but lightweight neural networks for
  real-time object detection}.
\newblock {\em arXiv preprint arXiv:1608.08021\/} (2016).
\newblock


\bibitem{krueger1985videoplace}
{Myron~W Krueger}, {Thomas Gionfriddo}, {and} {Katrin Hinrichsen}. 1985.
\newblock \showarticletitle{{VIDEOPLACE}---an artificial reality}. In {\em ACM
  SIGCHI Bulletin}, Vol.~16. 35--40.
\newblock


\bibitem{ladicky2014pulling}
{Lubor Ladicky}, {Jianbo Shi}, {and} {Marc Pollefeys}. 2014.
\newblock \showarticletitle{Pulling things out of perspective}. In {\em CVPR}.
  89--96.
\newblock


\bibitem{laina2016deeper}
{Iro Laina}, {Christian Rupprecht}, {Vasileios Belagiannis}, {Federico
  Tombari}, {and} {Nassir Navab}. 2016.
\newblock \showarticletitle{Deeper depth prediction with fully convolutional
  residual networks}. In {\em 3DV}. 239--248.
\newblock


\bibitem{leapmotion}
{LeapMotion}. [Online; accessed on 13-August-2019].
\newblock \url{https://www.leapmotion.com/}.
\newblock   ([Online; accessed on 13-August-2019]).
\newblock


\bibitem{malik2018deephps}
{Jameel Malik}, {Ahmed Elhayek}, {Fabrizio Nunnari}, {Kiran Varanasi}, {Kiarash
  Tamaddon}, {Alexis Heloir}, {and} {Didier Stricker}. 2018.
\newblock \showarticletitle{Deep{HPS}: End-to-end estimation of {3D} hand pose
  and shape by learning from synthetic depth}. In {\em 3DV}. 110--119.
\newblock


\bibitem{marquardt2011continuous}
{Nicolai Marquardt}, {Ricardo Jota}, {Saul Greenberg}, {and} {Joaquim~A Jorge}.
  2011.
\newblock \showarticletitle{The continuous interaction space: interaction
  techniques unifying touch and gesture on and above a digital surface}. In
  {\em IFIP Conference on Human-Computer Interaction}. 461--476.
\newblock


\bibitem{moser2016calibration}
{Kenneth~R Moser}, {Sujan Anreddy}, {and} {J~Edward Swan}. 2016.
\newblock \showarticletitle{Calibration and interaction in optical see-through
  augmented reality using leap motion}. In {\em IEEE VR}. 332--332.
\newblock


\bibitem{mueller2018ganerated}
{Franziska Mueller}, {Florian Bernard}, {Oleksandr Sotnychenko}, {Dushyant
  Mehta}, {Srinath Sridhar}, {Dan Casas}, {and} {Christian Theobalt}. 2018.
\newblock \showarticletitle{Ganerated hands for real-time {3D} hand tracking
  from monocular {RGB}}. In {\em CVPR}. 49--59.
\newblock


\bibitem{mueller2017real}
{Franziska Mueller}, {Dushyant Mehta}, {Oleksandr Sotnychenko}, {Srinath
  Sridhar}, {Dan Casas}, {and} {Christian Theobalt}. 2017.
\newblock \showarticletitle{Real-time hand tracking under occlusion from an
  egocentric {RGB}-{D} sensor}. In {\em ICCV}. 1284--1293.
\newblock


\bibitem{mur2015orb}
{Raul Mur-Artal}, {Jose Maria~Martinez Montiel}, {and} {Juan~D Tardos}. 2015.
\newblock \showarticletitle{{ORB-SLAM}: a versatile and accurate monocular
  {SLAM} system}.
\newblock {\em IEEE Transactions on Robotics\/} {31}, 5 (2015), 1147--1163.
\newblock


\bibitem{newell2016stacked}
{Alejandro Newell}, {Kaiyu Yang}, {and} {Jia Deng}. 2016.
\newblock \showarticletitle{Stacked hourglass networks for human pose
  estimation}. In {\em ECCV}. 483--499.
\newblock


\bibitem{nicodemou2018learning}
{Vassilis~C Nicodemou}, {Iason Oikonomidis}, {Georgios Tzimiropoulos}, {and}
  {Antonis Argyros}. 2018.
\newblock \showarticletitle{Learning to Infer the Depth Map of a Hand from its
  Color Image}.
\newblock {\em arXiv preprint arXiv:1812.02486\/} (2018).
\newblock


\bibitem{ono1988dynamic}
{Hiroshi Ono}, {Brian~J Rogers}, {Masao Ohmi}, {and} {Mika~E Ono}. 1988.
\newblock \showarticletitle{Dynamic occlusion and motion parallax in depth
  perception}.
\newblock {\em Perception\/} {17}, 2 (1988), 255--266.
\newblock


\bibitem{panteleris2018using}
{Paschalis Panteleris}, {Iason Oikonomidis}, {and} {Antonis Argyros}. 2018.
\newblock \showarticletitle{Using a single {RGB} frame for real time {3D} hand
  pose estimation in the wild}. In {\em WACV}. 436--445.
\newblock


\bibitem{radkowski2012interactive}
{Rafael Radkowski} {and} {Christian Stritzke}. 2012.
\newblock \showarticletitle{Interactive hand gesture-based assembly for
  augmented reality applications}. In {\em Proceedings of the 2012
  International Conference on Advances in Computer-Human Interactions}.
  303--308.
\newblock


\bibitem{ren2006figure}
{Xiaofeng Ren}, {Charless~C Fowlkes}, {and} {Jitendra Malik}. 2006.
\newblock \showarticletitle{Figure/ground assignment in natural images}. In
  {\em ECCV}. 614--627.
\newblock


\bibitem{rogez2015first}
{Gr{\'e}gory Rogez}, {James~S Supancic}, {and} {Deva Ramanan}. 2015.
\newblock \showarticletitle{First-person pose recognition using egocentric
  workspaces}. In {\em CVPR}. 4325--4333.
\newblock


\bibitem{saxena2006learning}
{Ashutosh Saxena}, {Sung~H Chung}, {and} {Andrew~Y Ng}. 2006.
\newblock \showarticletitle{Learning depth from single monocular images}. In
  {\em NIPS}. 1161--1168.
\newblock


\bibitem{saxena2008make3d}
{Ashutosh Saxena}, {Min Sun}, {and} {Andrew~Y Ng}. 2008.
\newblock \showarticletitle{{Make3D}: Learning {3D} scene structure from a
  single still image}.
\newblock {\em IEEE Transactions on Pattern Analysis and Machine
  Intelligence\/} {31}, 5 (2008), 824--840.
\newblock


\bibitem{song2015joint}
{Jie Song}, {Fabrizio Pece}, {G{\'a}bor S{\"o}r{\"o}s}, {Marion Koelle}, {and}
  {Otmar Hilliges}. 2015.
\newblock \showarticletitle{Joint estimation of {3D} hand position and gestures
  from monocular video for mobile interaction}. In {\em CHI}. 3657--3660.
\newblock


\bibitem{teo2015fast}
{Ching Teo}, {Cornelia Fermuller}, {and} {Yiannis Aloimonos}. 2015.
\newblock \showarticletitle{Fast {2D} border ownership assignment}. In {\em
  ICCV}. 5117--5125.
\newblock


\bibitem{google2018ar}
{Julien Valentin}, {Adarsh Kowdle}, {Jonathan~T. Barron}, {Neal Wadhwa}, {Max
  Dzitsiuk}, {Michael~John Schoenberg}, {Vivek Verma}, {Ambrus Csaszar},
  {Eric~Lee Turner}, {Ivan Dryanovski}, {Joao Afonso}, {Jose Pascoal},
  {Konstantine Nicholas~John Tsotsos}, {Mira~Angela Leung}, {Mirko Schmidt},
  {Onur~Gonen Guleryuz}, {Sameh Khamis}, {Vladimir Tankovich}, {Sean Fanello},
  {Shahram Izadi}, {and} {Christoph Rhemann}. 2018.
\newblock \showarticletitle{Depth from motion for smartphone AR}.
\newblock {\em ACM Transactions on Graphics (SIGGRAPH Asia)\/} {37}, 6 (2018),
  1--19.
\newblock


\bibitem{wang2018doobnet}
{Guoxia Wang}, {Xiaochuan Wang}, {Frederick~WB Li}, {and} {Xiaohui Liang}.
  2018b.
\newblock \showarticletitle{{DOOBN}et: Deep Object Occlusion Boundary Detection
  from an Image}. In {\em ACCV}. 686--702.
\newblock


\bibitem{wang2016doc}
{Peng Wang} {and} {Alan Yuille}. 2016.
\newblock \showarticletitle{{DOC}: Deep occlusion estimation from a single
  image}. In {\em ECCV}. 545--561.
\newblock


\bibitem{wang2009real}
{Robert~Y Wang} {and} {Jovan Popovi{\'c}}. 2009.
\newblock \showarticletitle{Real-time hand-tracking with a color glove}.
\newblock {\em ACM Transactions on Graphics (SIGGRAPH)\/} {28}, 3 (2009), 63.
\newblock


\bibitem{wang2018non}
{Xiaolong Wang}, {Ross Girshick}, {Abhinav Gupta}, {and} {Kaiming He}. 2018a.
\newblock \showarticletitle{Non-local neural networks}. In {\em CVPR}.
  7794--7803.
\newblock


\bibitem{ye2018occlusion}
{Qi Ye} {and} {Tae-Kyun Kim}. 2018.
\newblock \showarticletitle{Occlusion-aware hand pose estimation using
  hierarchical mixture density network}. In {\em ECCV}. 801--817.
\newblock


\bibitem{yuan2018depth}
{Shanxin Yuan}, {Guillermo Garcia-Hernando}, {Bj{\"o}rn Stenger}, {Gyeongsik
  Moon}, {Ju Yong~Chang}, {Kyoung Mu~Lee}, {Pavlo Molchanov}, {Jan Kautz},
  {Sina Honari}, {Liuhao Ge}, {and} {others}. 2018.
\newblock \showarticletitle{Depth-based {3D} hand pose estimation: From current
  achievements to future goals}. In {\em CVPR}. 2636--2645.
\newblock


\bibitem{zimmermann2017learning}
{Christian Zimmermann} {and} {Thomas Brox}. 2017.
\newblock \showarticletitle{Learning to estimate {3D} hand pose from single
  {RGB} images}. In {\em ICCV}. 4903--4911.
\newblock


\bibitem{Zimmermann2019ICCV}
{Christian Zimmermann}, {Duygu Ceylan}, {Jimei Yang}, {Bryan Russell}, {Max
  Argus}, {and} {Thomas Brox}. 2019.
\newblock \showarticletitle{Frei{HAND}: A Dataset for Markerless Capture of
  Hand Pose and Shape From Single {RGB} Images}. In {\em ICCV}.
\newblock


\end{thebibliography}

\end{document}